\documentclass[12pt]{article}
\usepackage{epsfig}
\usepackage{amssymb}

\def\Journal#1#2#3#4{{#1} {#2} (#4) #3 }
\def\NCA{{\em Nuovo Cimento} A}
\def\PHYS{{\em Physica}}
\def\NPA{{\em Nucl. Phys.} A}
\def\MATH{{\em J. Math. Phys.}}
\def\PRO{{\em Prog. Theor. Phys.}}
\def\NPB{{\em Nucl. Phys.} B}
\def\PLA{{\em Phys. Lett.} A}
\def\PLB{{\em Phys. Lett.} B}
\def\PLD{{\em Phys. Lett.} D}
\def\PL{{\em Phys. Lett.}}
\def\PRL{\em Phys. Rev. Lett.}
\def\PREV{\em Phys. Rev.}
\def\PREP{\em Phys. Rep.}
\def\PRA{{\em Phys. Rev.} A}
\def\PRD{{\em Phys. Rev.} D}
\def\PRC{{\em Phys. Rev.} C}
\def\PRB{{\em Phys. Rev.} B}
\def\ZPC{{\em Z. Phys.} C}
\def\ZPA{{\em Z. Phys.} A}
\def\ANNP{\em Ann. Phys. (N.Y.)}
\def\RMP{{\em Rev. Mod. Phys.}}
\def\CHEM{{\em J. Chem. Phys.}}
\def\INT{{\em Int. J. Mod. Phys.} E}

\newcommand{\be}{\begin{equation}}
\newcommand{\ee}{\end{equation}}
\newcommand{\bea}{\begin{eqnarray}}
\newcommand{\eea}{\end{eqnarray}}

\begin{document}
\vspace*{0.5cm}
\begin{center}
{\bf SPIN PHYSICS AT COMPASS~\footnote{
invited talk at the International School of Nuclear Physics, 26-th Course: Lepton scattering 
and the structure of hadrons and nuclei, Erice, 16 - 24 September 2004}}\\
\vspace*{1cm}
F. Bradamante\\
\vspace*{0.3cm}
Dipartimento di Fisica, Universit\`a di Trieste,
and INFN, Sezione di Trieste, Italy\\
\vspace*{1cm}
on behalf of the COMPASS Collaboration
\end{center}
\vspace*{1cm}

\begin{abstract}
COMPASS is a new fixed target experiment presently in operation at CERN.
It has the goal to investigate hadron
structure and hadron spectroscopy by using either muon or hadron beams.
From measurements of various hadron asymmetries in polarized muon - nucleon
scattering it will be possible to determine the
contribution of the gluons to the nucleon spin.
Main objective of the hadron program is the search of exotic states, and
glueballs in particular.
This physics programme is carried out with a two-stage magnetic
spectrometer, with particle identification and calorimetry in both
stages, which has started collecting physics data in 2002, and will
run at the CERN SPS at least until 2010.
Preliminary results from the 2002 run with a 160 GeV muon beam are presented for several physics channels under
investigation.
\end{abstract}

\vskip 8mm
\section{Introduction}
\label{sec:xxx}
Five years ago I had the pleasure to illustrate to the participants to the
21-st course of the International School of Nuclear Physics the concept
and the physics goals of the COMPASS experiment.
In 1996 an international collaboration comprising some 200 physicists
from 30 Institutes proposed COMPASS~\cite{cprop}, a COmmon Muon and Proton Apparatus for
Structure and Spectroscopy, a new state-of-the-art two-stage fixed
target magnetic spectrometer to run at CERN and carry on an ambitious 
research programme aiming at a deeper understanding of nucleon
structure and confinement. 
The experiment was approved in 1997, and the bulk of the apparatus constructed over the
following 4-5 years.

The main physics observables studied
are the polarization of the constituents of a polarized
nucleon, the mass and decay patterns of light hadronic systems
with either exotic quantum numbers or strong gluonic
excitations, and the leptonic decays of charmed hadrons.

A possible polarization of gluons $\Delta G / G$ 
in a polarized nucleon is searched for by the study of hard
processes in  polarized muon -- polarized nucleon
deep inelastic scattering (DIS),
open charm production
and high $p_T$--meson pair production. 
Using very large event
samples COMPASS should determine
for the first time $\Delta G/G$ in the
kinematical
region of $x_{gluon}$ between 0.05 and 0.3. 
The flavour-separated spin
distribution functions of the nucleon in 
deep inelastic scattering are also being measured, both in longitudinal
and transverse polarization mode. 
In the latter the still unmeasured
transversity distribution $\Delta_T q$ will be investigated.

With a hadron beam,
gluonic degrees of freedom shall be excited in hadrons using diffractive
and double-diffractive scattering.
High statistics measurements will allow to access the mass range above
2 GeV/$c^2$.
Leptonic and semi-leptonic decays of charmed hadrons will be studied using a
specialized detector arrangement
to identify such processes and discriminate background.
In addition many soft processes can be studied testing low energy theorems
of QCD.

What I had anticipated five years ago is now a reality.
After a long and hard technical run in 2001, the experiment 
started taking data with the
muon beam in 2002.
Data taking was continued in 2003 and in 2004, again with the muon beam,
and the 2004 run is presently still ongoing.
To pave the way to the hadron beam measurements, a 
3-week pilot run with a hadron beam is planned
at the end of the 2004 run.

After the technical stop of all the accelerators at 
CERN in 2005, the experiment 
will resume data taking and is expected to be in operation
until at least 2010.

\section{The COMPASS spectrometer}
The COMPASS spectrometer has been set up at the CERN SPS 
muon beam, in Hall 888 of the Prevessin site. 
It combines  high rate beams with a modern two
stage fixed target magnetic spectrometer. 
Both stages foresee charged particle identification with fast 
RICH detectors,
electromagnetic calorimetry, hadronic
calorimetry, and muon identification via filtering through thick absorbers.
 The design of detector
components, electronics and data acquisition system allows to
handle beam rates up to 10$^8$ muons/s and about 5$\cdot$10$^7$
hadrons/s with a maximal interaction rate of about
2$\cdot$10$^6$/s. 
The triggering system and the tracking system of COMPASS have been designed to
stand the associated rate of secondaries, and use state-of-the-art detectors.
Also, fast front-end electronics, multi-buffering, and a large
and fast storage of events are essential.

\begin{figure*}[hbt] %
\centerline{\epsfig{file=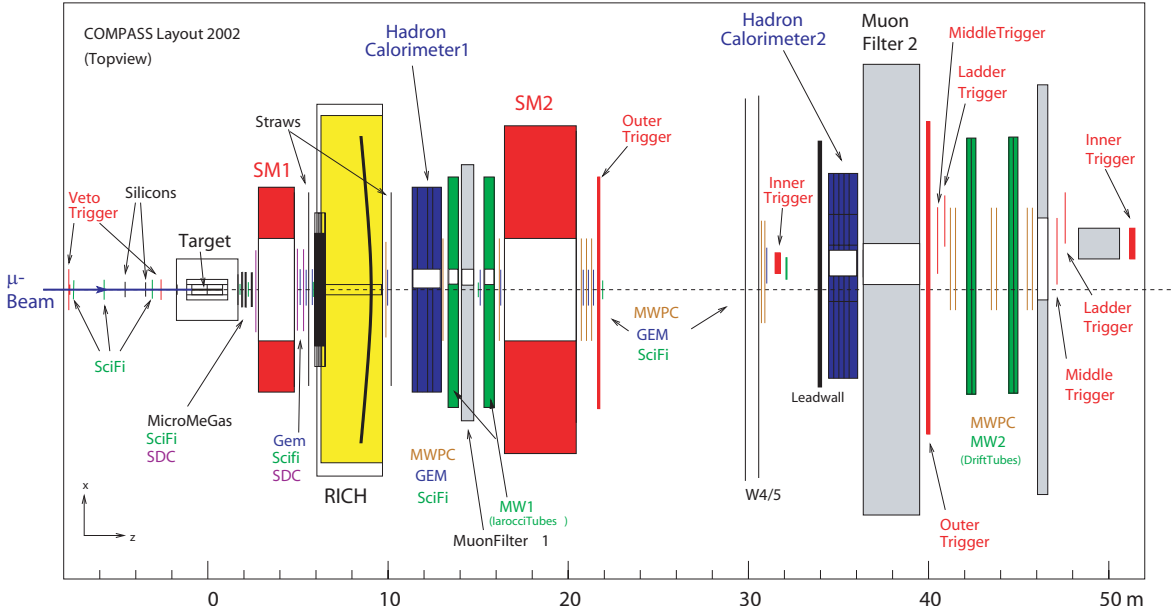,width=155.mm}}
\caption{Top view of the lay-out of the spectrometer for 
the COMPASS experiment in 2002.
The labels and the arrows refer to the major components of the
tracking, trigger, and PID systems.}
\label{fig:setup}
\end{figure*}

The layout of the spectrometer which was on the floor in 2002 is shown 
in Fig.~\ref{fig:setup}.

The experiment has been run at a muon energy of 160 GeV.
The beam is naturally polarized by the $\pi -$decay mechanism,
and the beam polarization is estimated to be 76\%.
The  beam intensity is $2\cdot 10^8$ muons per spill (4.5 s long).

We use the polarized target system of the SMC experiment,
which allows for two oppositely polarized target cells,
60~cm long each.
The PT magnet can provide both a solenoidal field (2.5 T) and a dipole field
(0.5 T), for adiabatic spin rotation and for the transversity measurements.
Correspondingly, the
target polarization can then be oriented either longitudinally
or transversely to the beam direction.
Use of two different target materials, NH$_3$ as proton target and
$^6$LiD as deuteron target, is foreseen. 
Polarizations of 85~\% and 50~\% have been reached, respectively.
In so far we have used $^6$LiD: its favourable dilution factor of $\sim$0.5
is of the utmost importance for the measurement
of $\Delta G$.

To match the expected particle flux in the various locations
along the spectrometer, COMPASS uses very different tracking detectors.
The small area trackers consist of several
stations of scintillating fibres, silicon detectors, micromegas
chambers~\cite{microo} and gas chambers using the GEM-technique~\cite{gems}.
Large area tracking devices are made from gaseous detectors
(Saclay Drift Chambers, Straw tubes~\cite{straw}, MWPC's, 
and W4/5 Drift Chambers) placed around the two spectrometer magnets. 
Table~\ref{tab:res02} summarizes the spatial resolution and the timing 
properties of the tracking detectors, as derived from the 2002 data.
\begin{table}[htb]
\begin{center}
\begin{tabular}{|l|c|c|c|c|}
\hline
Detector & number of & efficiency & resolution & timing \\
         &  coordinates &  &  &  \\
\hline
Scintillating fibers  & 21 & 94 \%       & 130 $\mu$m  & 0.45 ns \\
Micromegas            & 12 & 95 - 98 \%  &  65 $\mu$m  & 8 ns \\
GEM                   & 40 & 95 - 98 \%  &  50 $\mu$m  & 12 ns \\
SDC                   & 24 & 94 - 97 \%  & 170 $\mu$m  &  \\
Straw tubes           & 18 & $>$ 90 \%     & ~270 $\mu$m &  \\
MWPC                  & 32 & 97 - 99 \%  &             &  \\
W4/5                  &  8 & $>$ 80 \%     &             &  \\
\hline
\end{tabular}
\caption{Trackers performances in the 2002 run.}
\label{tab:res02}
\end{center}
\end{table}
Muons are identified in large-area Iarocci-like tubes and drift 
tubes downstream of muon absorbers.

The charged particle identification relies on the RICH technology.
Presently, only RICH1 (the RICH in the first magnetic spectrometer)  exists~\cite{rich1}.
The length of the radiator (C$_4$F$_{10}$ gas) vessel is 3~m.
The entire downstream surface is covered by 116 aluminized
mirrors with spherical geometry and a focal length of 3.3~m.
As VUV photon detectors we use MWPC's with a CsI photocathode~\cite{RD26} 
(segmented in $8\times8$~mm$^2$ pads) which detect photons with wave length 
shorter than 200 nm, i.e. in the far UV domain.
The active area of each of the two photon detectors is 2.8~m$^2$
and the total number of pads is about 70,000.
The front-end electronics uses a modified version of the Gassiplex chip,
and the read-out cards constitute
a major project, utilizing hundreds of DSP's.

The trigger is formed by two hadron 
calorimeters and several hodoscope systems.
Electromagnetic calorimetry is being installed upstream of the hadronic
calorimeters at the end of each spectrometer section. 

The readout system~\cite{daq} uses a modern concept, involving highly 
specialized integrated circuits. 
The readout chips are placed close to the detectors and the data are 
concentrated at a very early stage via high speed serial links. At the 
next level high bandwidth optical links transport the data to a system 
of readout buffers. 
The event building system is based on PCs and Gigabit 
or Fast Ethernet switches and is highly scalable.
This high performance network is also used to transfer the assembled data 
to the computer center for database formatting, reconstruction, analysis 
and mass storage.
The data are sent via an optical link from the Hall 888 
directly to the Computer building for Central Data Recording
(CDR).

To handle the huge amount of data (the collected raw data size is 
$\sim$300 TB/year)
we used Objectivity/DB
until the end of 2002~\cite{off-line}, and Oracle since.
The  power needed to process COMPASS data is about 100 kSI2k.
In the off-line farm, the data servers handle the network traffic
from the CDR, distribute the raw data to the CPU 
clients (where they are put in the data base), receive them back from
the PCs, and finally send them to a hierarchical storage manager (HSM) 
system.
In parallel, the data servers receive the data to be processed from the
HSM, send them to the PCs for processing, collect the output (DST or mDST),
and send it to the HSM.
Data processing is performed on the farm at CERN while DST
data analysis is done on satellite farms in the major home institutes.

A major effort was devoted to write from scratch the off-line
programs (CORAL, the new COmpass Reconstruction and AnaLysis program)
using object-oriented technology and C++ language. 

\section{First physics results}
\label{sec:data}
Over the 80 days of the 2002 run, a total of about 6000 millions of events 
have been collected, corresponding to a data size of 260 TB.
Similar numbers have been taken in the 2003 run and are presently being
collected in the 2004 run.
Many different physics topics are presently being investigated,\\
- $\Delta G$ from open charm and high $p_T$ hadron pair,\\
- $A_{LL}$ to extract $A_1^d$, the virtual photon asymmetry, and $g_1$,\\
- vector meson ($\rho, \; \phi, \; J/\Psi$) production to 
test s-channel helicity conservation,\\
- $\Lambda$ physics,\\
- transversity (single hadron, hadron pairs, $\Lambda$),\\
- Cahn asymmetries,\\
- search of exotics ($\Theta^+, \; \Xi^{--}, \; ...$).

A flavour of the physics results obtained from the 2002 data is given below.

\subsection{$\Delta G / G$ from charm photo-production}

\begin{figure*}[htb] %
\centerline{\includegraphics[width=2.3in]{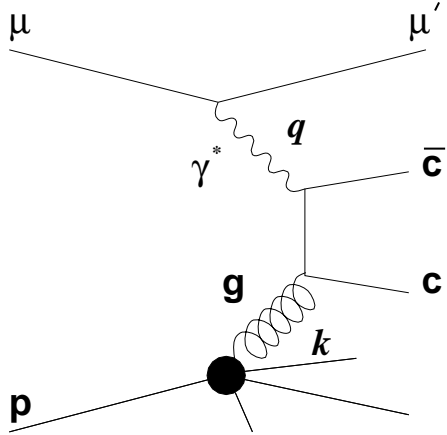}}
\caption{The photon-gluon fusion diagram,
dominant mechanism for charm production at COMPASS energies.}
\label{fig:diag}
\end{figure*}

The main goal of COMPASS is a direct measurement of $\Delta G$
by measuring the cross-section asymmetry  of open charm in DIS 
$A_{\mu N}^{c \bar c}$
$$
A_{\mu N}^{c \bar c} = \frac{\Delta  \sigma^{\mu N\rightarrow 
c\bar{c}X}}{ \sigma^{\mu N\rightarrow c\bar{c}X}} .
$$
The possibility to measure directly  $\Delta G$ by measuring
$A_{\mu N}^{c \bar c}$ was put forward already sixteen years ago~\cite{Altarelli88,Gluck88}
to solve the nucleon spin puzzle raised by the EMC result~\cite{EMC88}.
At COMPASS energies the production of charm goes predominantly
via photon-gluon fusion (PGF), according to the diagram shown in 
Fig.~\ref{fig:diag}, and the quantities $\sigma^{\mu N\rightarrow c\bar{c}X}$
and $\Delta  \sigma^{\mu N\rightarrow c\bar{c}X}$ can be expressed as a 
convolution of the elementary photon-gluon cross-section with the gluon 
distributions $G$ and $\Delta G$.

\begin{figure}[hbt] %
\centerline{\includegraphics[width=3.3in]{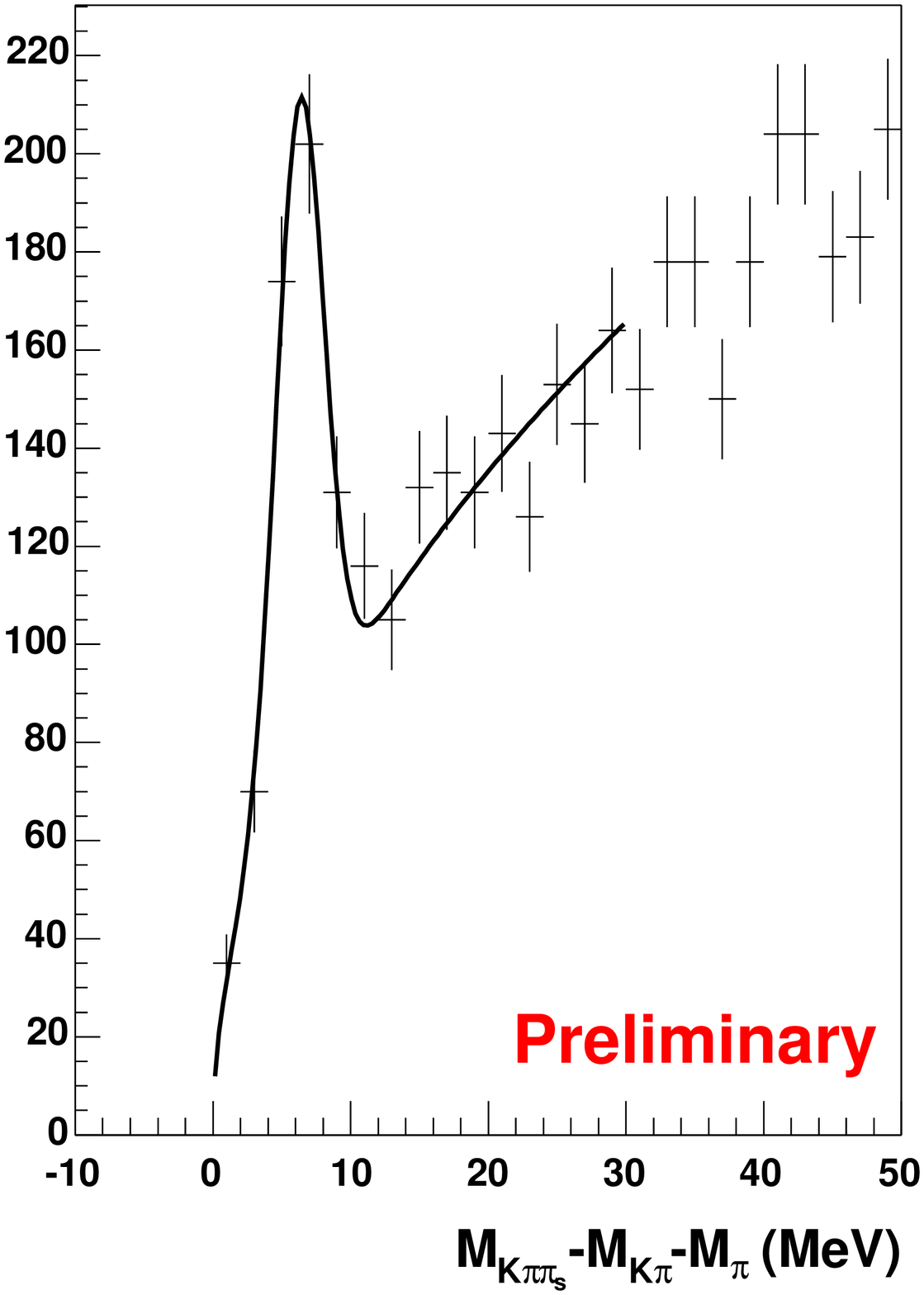}
\includegraphics[width=3.3in]{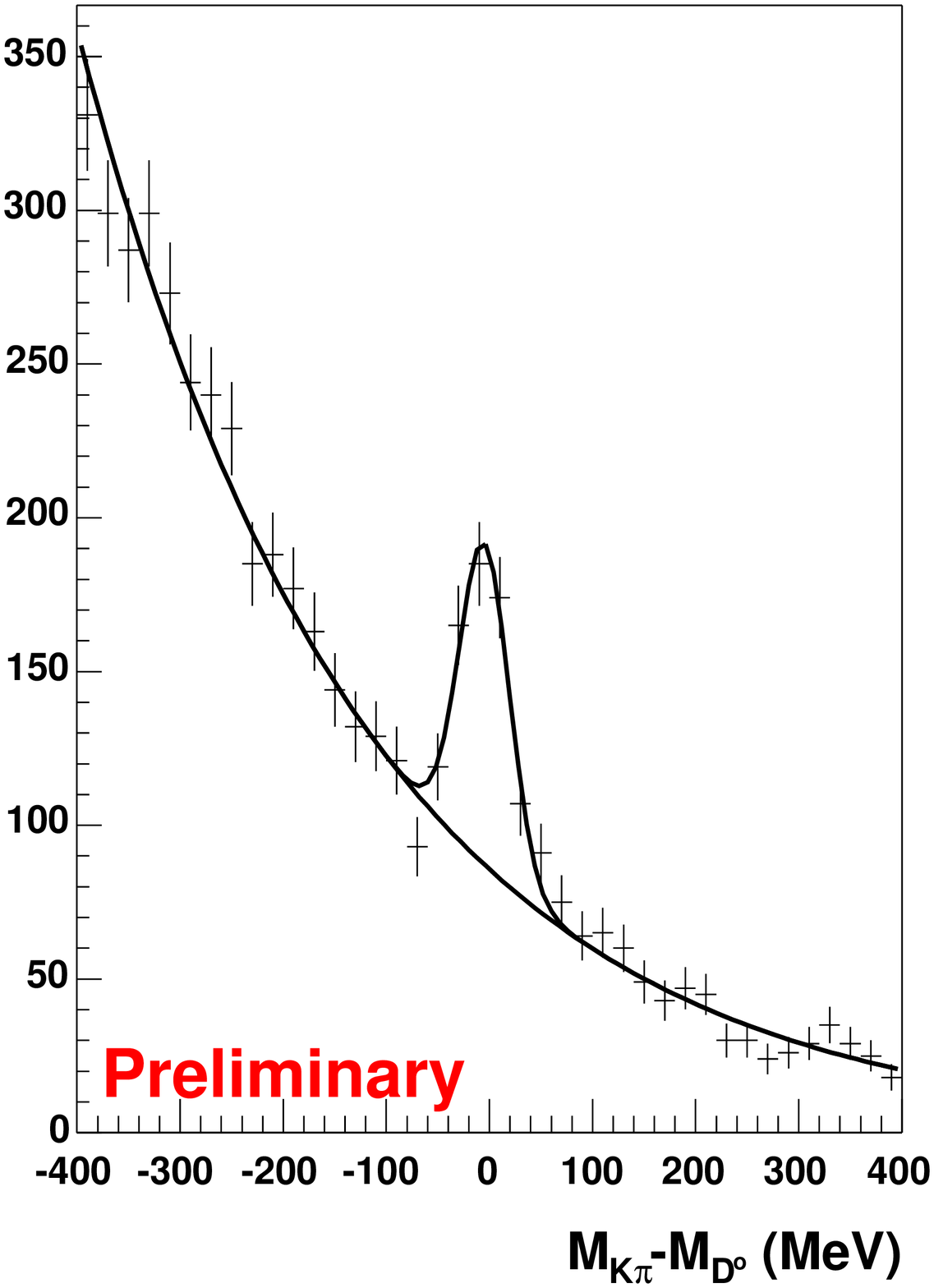}}
\vspace*{-0.3cm}
\caption{
Left:
$D^*$  produced by requiring the invariant 
mass of the $K\pi$  pair to be in a 60 MeV 
window around the
$D^0$ peak and a soft pion being detected. 
Selecting the events around the $D^*$  mass,
the $D^0$  peak in the 
invariant mass spectrum of $K\pi$ is very clear
(figure at the right).}
\label{fig:op1}
\end{figure}

Open-charm events are identified by reconstructing
$D^\circ{}$, $\bar{D}^\circ{}$, and $D^{*\pm}$ mesons from their decay 
products, i.e. 
$D^\circ{} \rightarrow  K^-\pi^+$ and 
$D^{*+} \rightarrow  D^\circ{} \pi^{+} \rightarrow K^-\pi^+ \pi^+$
and charge conjugate.
In the first case, cuts on the $K$ direction in the
$D^\circ{}$ rest frame ($|cos(\theta^*_K)| < 0.5$) and on the $D^\circ{}$
  energy fraction 
($z_D = E_D/E_{\gamma^*} > 0.25$) are needed to reduce the background 
contamination. 
Preliminary signals of the D mesons are shown in fig.~\ref{fig:op1}.
Kaon-pion pairs are selected by asking:
$z_D > 0.2$; $|cos(\theta^*_K)| < 0.85$; $10 < p_K < 35$ GeV in 
order to be 
in the RICH $K$ identification region. A soft pion ($< 10$ GeV) 
is also required.
This measurement is statistically limited.
The $D^\circ{}$ signal in our data is at the level of 10$^{-7}$,
and other decay channels are also being investigated presently.
From the present analysis and this channel only the projected error on
$\Delta G / G$ using the data from 2002, 2003 and 2004 should be
about 0.24.

\subsection{$\Delta G / G$ from high $p_T$ hadron pairs}

The most promising additional 
way to measure $ \Delta G$ in COMPASS uses the asymmetry
of charged hadron pairs at high $p_T$~\cite{bravar}.
Originally developed for the COMPASS experiment,
the method has been recently applied also to the HERMES data~\cite{hermespt}.
The basic diagram is still the PGF, 
$\gamma g \rightarrow q \bar{ q} \rightarrow h^+ h^- X$,
and the hardness of the process is guaranteed by the large $p_T$.
The background from the leading order process 
$\gamma q \rightarrow q$, and the QCD-Compton process,
$\gamma q \rightarrow \gamma q$, is in general dominating the PGF
creation of a light $q\bar{q}$ pair, but suitable kinematic cuts
can enhance considerably this process and allow a statistically
precise measurement.

The  virtual photon-deuteron asymmetry $A^{\gamma* d}$ for high $p_T$  
pair production 
evaluated from the 2002 data for all $Q^2$, is found to be  
$A^{\gamma* d} = -0.065\pm0.036(stat) \pm0.010(syst.)$,
a hint that $\Delta G/G$ could be small and positive. 
The systematic error is estimated only from measurements of
the so-called ``false asymmetries'', i.e. asymmetries which should be
zero for an apparatus with uniform acceptance.
Prior to calculation of $\Delta G/G$ from $A^{\gamma* d}$, one needs 
to subtract the asymmetries from the physics background.
Physics background contaminations can be reduced by selecting events 
with $Q^2 >1$ GeV$^2$, but this cut substantially reduces statistics. 
Assuming that the contribution of 
PGF processes to the measured asymmetry is about 1/4 of the 
total and using all 
statistics expected at COMPASS by the end of 2004 run, one 
can determine $\Delta G/G$
with an accuracy $\sigma(\Delta G/G) \simeq$ 0.05 for all 
$Q^2$ and $\simeq$~0.17 for events with 
$Q^2 >1$ GeV$^2$.

\subsection{Virtual photon-deuteron asymmetry $A_1^d(x)$}

\begin{figure}[hbt] %
\centerline{\includegraphics[width=5.in]{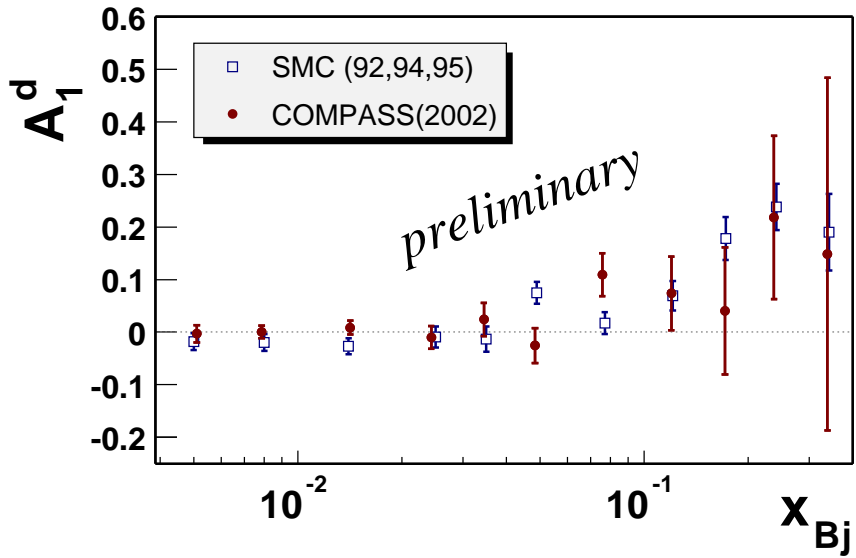}
}
\vspace*{-0.3cm}
\caption{
Virtual Photon-Deuteron asymmetry as a function of $x$-Bjorken compared 
to the SMC published data.}
\label{fig:a1d}
\end{figure}

This asymmetry is related to the polarized structure function  and 
has been measured 
before  by SMC~\cite{smc98}, SLAC~\cite{slac98} and HERMES~\cite{hermesvp} 
collaborations. Potential opportunities of 
COMPASS, especially in the region of small $x$, are seen from the preliminary 
COMPASS 
data  on  $A_1^d(x)$ obtained in 2002 (Fig.~\ref{fig:a1d}). 
This sample includes about 6.5$\cdot 10^6$ events within usual DIS cuts
($Q^2 >1$ GeV$^2$). 
Also shown in figure~\ref{fig:a1d} are the SMC data which are the only ones to 
also cover the low-$x$ region.
The two sets of data agree rather well although the new COMPASS data do not 
hint at the slightly negative values of $A_1^d$ which were suggested by the 
SMC data.
The statistical errors of the COMPASS data in the low-$x$ region 
($x < 3\cdot 10^{-2}$)
are already now smaller than the SMC ones, and will decrease by a factor
of two when the data from 2003 and 2004 will be added to the analyzed sample.
The final accuracy we will obtain in the low $x$ region 
will allow to improve the measurement of the first moment of $g_1$, and 
thus $\Delta \Sigma$.

The addition of a new large-$Q^2$ trigger system in 2003 has already 
resulted in data
points at large $x$ with an accuracy comparable to that of
the SMC experiment,
therefore by 2004 even the large-$x$ data of COMPASS will be more 
precise than the
SMC data.

\subsection{Collins asymmetry and transversity}
As originally shown by Jaffe and Ji~\cite{Jaffe91}, to completely
specify the quark state at the twist-two level one has to add 
the transverse spin distributions $\Delta_T q(x)$ to the momentum
distribution $q(x)$ and to the helicity distribution $\Delta q(x)$.
Measurement of $\Delta_T q(x)$ gives access to new information
related to relativistic effects for bound quark states, the study of new 
evolution in QCD, the knowledge of the tensor
charge of the nucleon, and predictions for other processes 
involving transversity.

The transversity distributions $\Delta_T q(x)$ have never been measured, 
since they
are chirally-odd functions and do not contribute to
inclusive deep inelastic scattering. They  may instead be
extracted from measurements of the spin asymmetries in cross-sections 
for semi-inclusive deep inelastic scattering (SIDIS)
between leptons and transversely polarized nucleons, in which 
a hadron is also detected in the final state. In
such processes the measurable asymmetry is due to the combined effect 
of $\Delta_T q(x)$ and another chirally-odd function, which
contributes to the fragmentation of the transversely polarized quark. 
In the case in which the observed final hadron
is a pion or, in general, a scalar particle, this new fragmentation 
function is the so-called Collins function $\Delta_T D_q^h$
which describes the hadronization of a transversely polarized quark $q$ in
a hadron $h$~\cite{Collins93}, as yet
unmeasured, which in its own right merits serious study. 
In the case in
which the  observed final particle is for
example a $\Lambda^\circ{}$, the chirally-odd function is a transverse 
fragmentation 
function, also unknown and interesting. Other
channels for accessing $\Delta_T q$  require the detection of a 
vector particle 
or two pions in the final state. 

\begin{figure}[hbt]
\hspace*{0.35cm}
\begin{center}
\vspace*{-1.3cm}
\includegraphics[width=5.5in]{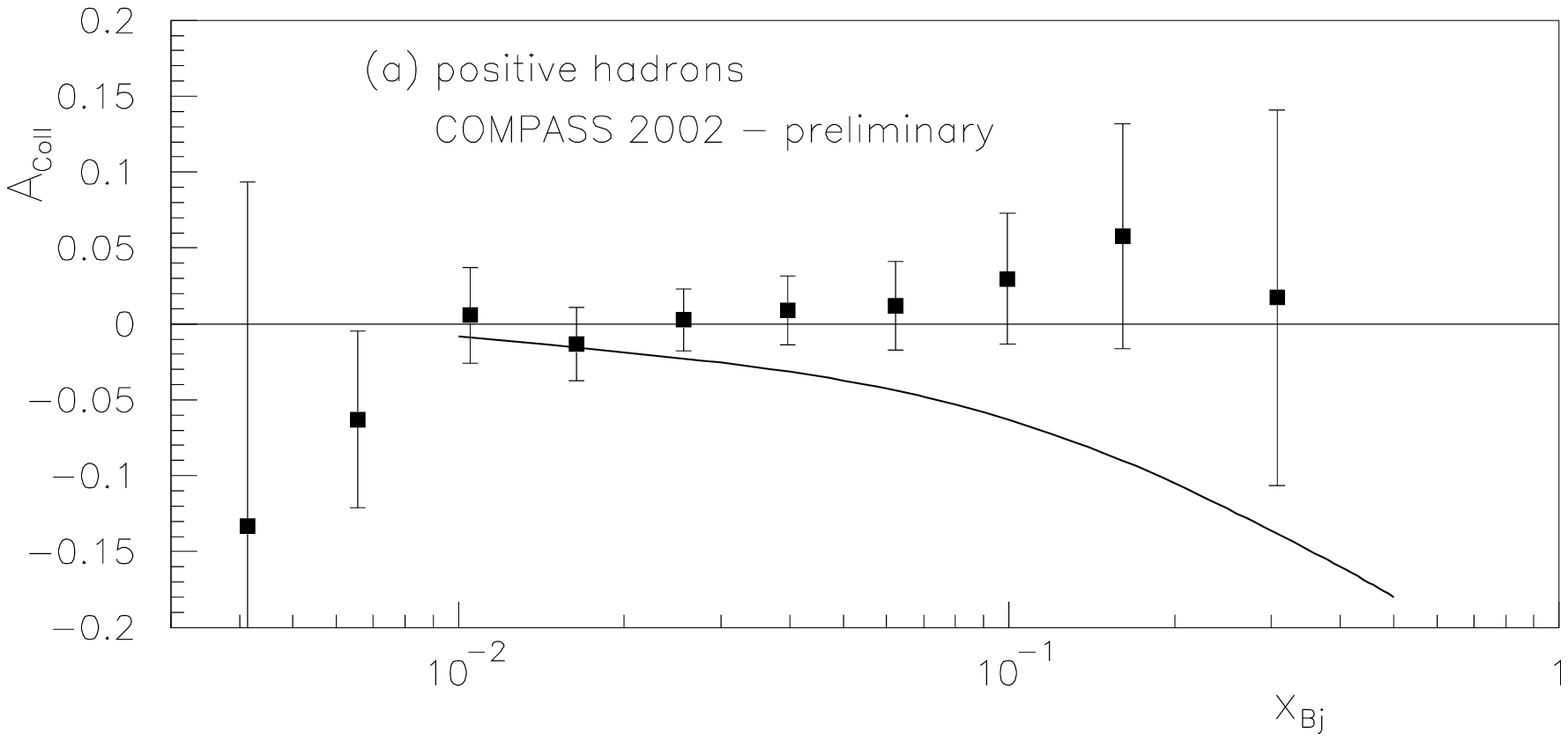}\\
\vspace*{-7.8cm}
\includegraphics[width=5.5in]{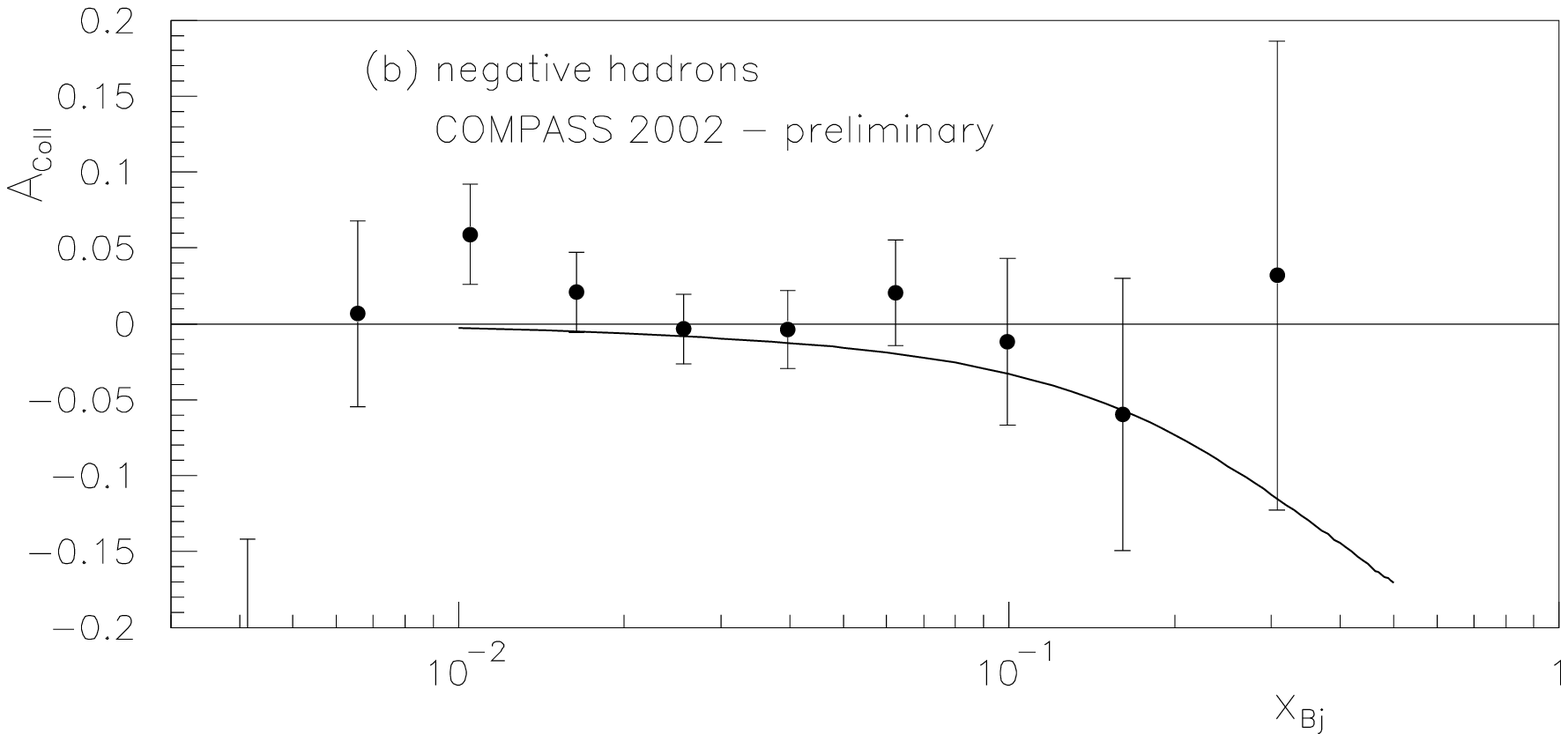}\\
\vspace*{-7.0cm}
\caption{Preliminary results for the asymmetry $A_{UT}$ 
for positive (a) and negative (b) leading hadrons produced by 
160 GeV $\mu^+$ on a transversely polarized deuterium target.
The data refer to the year 2002 COMPASS run, while the curves are
from Ref.~\cite{efremov}.}
\label{fig:acoll}
\end{center}
\end{figure}

To measure transversity, COMPASS has taken SIDIS data with the
$^6$LiD polarized along the vertical direction, i.e. orthogonally
to the incoming $\mu$ momentum.
About 20\% of the COMPASS running time has been devoted
to this measurement. 
Figure~\ref{fig:acoll} shows preliminary values of the first ever measured
single hadron Collins asymmetry on a deuteron target, separately for
positive and  negative leading hadrons.
The data are compared with a model calculation~\cite{efremov} 
of the asymmetry $A_{UT}$, which includes the transversity distribution
function $\Delta_T q(x)$  through the
following linear combination of quark flavours:
$$
A_{UT}(x) = \frac{\Sigma e_q^2 \cdot \Delta_T q(x)\cdot \Delta_T D_q^h}
{\Sigma e_q^2 \cdot  q(x)\cdot D_q^h} .
$$
The small values of the measured asymmetries at all $x$ might imply
either a cancellation between the proton and the neutron asymmetries,
or a small Collins effect in the fragmentation.

Parallel work on the Collins asymmetry of hadron pairs, and on the transverse
polarization of $\Lambda$'s produced in the transversity runs,
is ongoing.

\subsection{Exclusive $\rho^0$ production}

The large amount of data taken by COMPASS over a large kinematical domain make
the investigation of several dynamical processes possible.
A good example is provided by the study we have performed of the
exclusive production of the $\rho^0$ vector meson on the nucleon 
$\mu + N \rightarrow \mu' +N' + \rho^0$.
It is known from previous experimental data that the helicity of
the photon in the $\gamma* N$ center-of-mass system is approximately
retained by the vector meson, a phenomenon known as
s-channel helicity conservation (SCHC).
An accurate measurement of the violation of the SCHC will give
 further insight into the interaction.
Using our longitudinally polarized muon beam we could measure the full set
of $\rho^0$ spin density matrix elements~\cite{frho0} and determine 
their $Q^2$ dependence.
From the 2002 run the total sample of good accepted events,
with $Q^2 > 0.01$ GeV$^2$ is about 700000.

\begin{figure}[hbt]
\centerline{\includegraphics[width=4.5in]{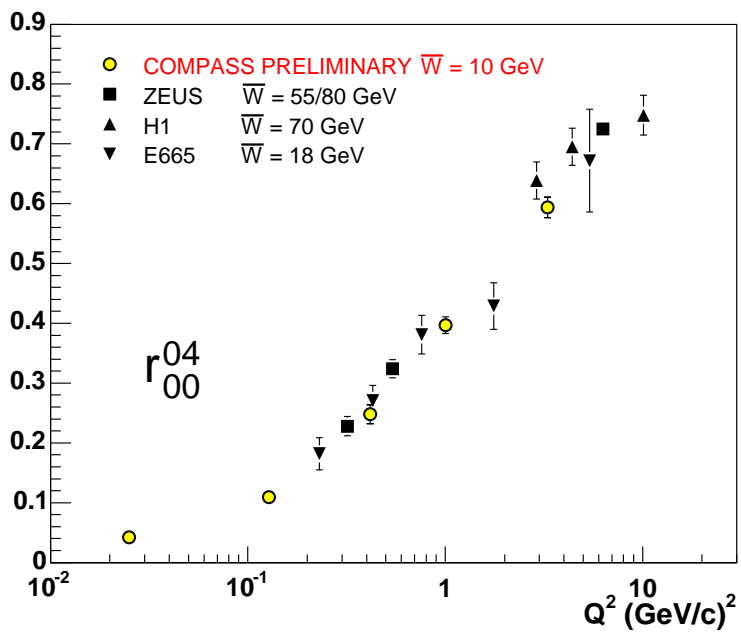}}
\caption{Preliminary results for the $Q^2$ dependence of the $\rho^0$ 
spin density
matrix element $r_{00}^{04}$ of the $\rho^0$ exclusive production
in $\mu N$ scattering. The COMPASS data are compared with existing 
measurements.}
\label{fig:rho0}
\end{figure}

From the decay angular distribution of the two pions in the $\rho^0$
center-of-mass system we could determine the element
$r_{00}^{04}$.
Our result for this element, which can be interpreted as 
the fraction of longitudinal (helicity 0) $\rho^0$ in the sample,
are given in fig.~\ref{fig:rho0}, where they are also compared with
the results of other experiments~\cite{zeus,h1,e665}.
This result indicates that at small $Q^2$ the production by
transverse photons almost completely dominates.

\begin{figure}[hbt]
\centerline{\includegraphics[width=5.2in]{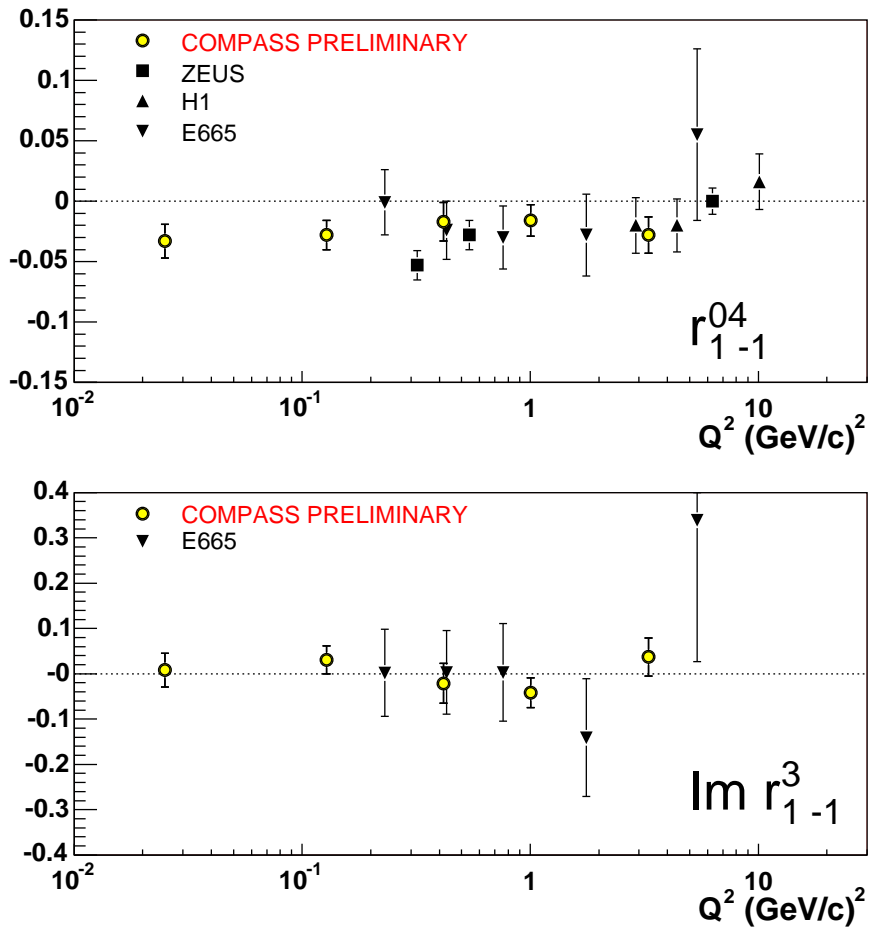}}
\caption{Preliminary results for the $Q^2$ dependence of the $\rho^0$ 
spin density
matrix elements $r_{1-1}^{04}$ and $Im \, r_{1-1}^{3}$
in  $\rho^0$ exclusive production
in $\mu N$ scattering. The COMPASS data are compared with existing 
measurements.}
\label{fig:rho0a}
\end{figure}

From the distribution of the angle between the $\rho^0$
production plane and  the $\rho^0$ decay plane we could determine the
two matrix elements $r_{1-1}^{04}$ and $Im \, r_{1-1}^{3}$.
It has to be stressed that the latter one can only be 
determined using a polarized
lepton beam.
As shown in fig.~\ref{fig:rho0a} our measurements exhibit
small negative values of $r_{1-1}^{04}$ ($\approx$ -0.03),
approximately independent of $Q^2$, whereas $Im \, r_{1-1}^{3}$
is consistent with zero.
The non-zero value of $r_{1-1}^{04}$ indicates
a small contribution of amplitudes with helicity-flip,
i.e. a small SCHC violation.
Again our data are compared in the figure with the data from other
experiments~\cite{zeus,h1,e665}.
%

\section{Conclusions}
It has been a pleasure to come to Erice and to present the COMPASS experiment 
at this most interesting School, and I am grateful to Amand for all the
work and enthusiasm he puts in this initiative.

The COMPASS experiment is an important effort to progress in the
understanding of the material world in which we live.
It was born in 1996, when times were really hard at CERN.
It has taken a very large effort to build it, but now it is a 
running experiment.

    The experiment is presently half-way in its third year of
data taking, and
is expected to run for several more years. I have described the
variety of detectors we are using and given numbers for their characteristic 
responses. For several of these detectors, this is the first time
they are used in an experiment, the necessary R\&D work having been done
for them by groups of COMPASS physicists. I have shown first physics
results from the data analysis, and underlined the very large effort which
is presently ongoing to fully understand the spectrometer and the
data. 
From these first results it should be clear that COMPASS is fulfilling
its promises and is already showing its huge physics potential.
As an example I have mentioned the status and the preliminary results of a few
analysis, $A_{LL}, \; \Delta G,$ exclusive $\rho^0$ production,
and transversity, from the 2002 data set.
Many more results based on 2002 and 2003 data will be presented
at SPIN2004~\cite{spin2004}.


\end{document}